\documentclass[12pt]{article}
\usepackage{graphicx}
\usepackage[totalwidth=480pt, totalheight=680pt]{geometry}

\title{Micro Pixel Chamber with resistive electrodes\\ for spark reduction}

\author{Atsuhiko Ochi$^a$\thanks{Corresponding author.}, 
Yuki Edo$^a$, Yasuhiro Homma$^a$,\\
 Hidetoshi Komai$^a$~and
Takahiro Yamaguchi$^a$ \\
\llap{$^a$}Kobe University\\
Kobe 657-8501, Japan\\
E-mail:  ochi@kobe-u.ac.jp }

\begin{document}

\maketitle

\abstract{The Micro Pixel Chamber ($\mu$-PIC) using resistive electrodes
has been developed and tested. 
The surface cathodes are made from resistive material, 
by which the electrical field is reduced when large current is flowed.
Two-dimensional readouts are achieved by anodes and pickup electrodes,
on which signals are induced. 
High gas gain (~$> 6\times 10^4$~) was measured using $^{55}$Fe (5.9 keV) source,
and very intensive spark reduction was attained under fast neutron.
The spark rate of resistive $\mu$-PIC was only $10^{-4}$ times less than
that of conventional $\mu$-PIC at the gain of $10^4$.
With these developments, a new MPGD with no floating structure is achieved,
with enough properties of both high gain and good stability
to detect MIP particles.
In addition, $\mu$-PIC can be operated with no HV applied on anodes
by using resistive cathodes. 
Neither AC coupling capacitors nor HV pull up resisters are needed
for any anode electrode.
Signal readout is drastically simplified by that configuration. }

\vspace{10pt}
\noindent {\bf  keywords:} Gaseous detector; Micropattern gaseous detectors (MSGC, GEM, THGEM, RETHGEM, MHSP, MICROPIC, MICROMEGAS, InGrid, etc), Micro pixel chamber

\section{Introduction}
The $\mu$-PIC is one type of micro pattern gaseous detector (MPGD)
\cite{upic_1, upic_2}. There is a dot of anode and a surrounded cathode for 
one pixel, and many of those are printed on substrate. 
Primary electrons in gas volume are drifted toward the anode, 
and those are multiplied by a gas avalanche mechanism due to higher electric field 
between anode and cathode. 
The $\mu$-PIC can be operated around $10^4$ of gas multiplication, 
which enables us to detect the minimum ionized particle. 
However, if there are large energy deposits simultaneously, such as recoiled nuclei, 
discharge and spark probabilities are raised because of dense electrons
called the "Raether limit" \cite{Raether_0, Raether_1, Raether_2}.
Those are a critical problem for using MPGD in intense hadron experiments. 
The most promising strategy for overcoming the spark problem
is to make the MPGDs using high resistive electrodes \cite{resistive_1, resistive_2}.
In the early stages of our investigation,
we tried to overcoat high resistivity materials on the cathode electrodes
\cite{tipp2011_ochi, mpgd2011_ochi}.
However, the sparks could not be suppressed by those early prototypes.
With the improvement of structure, separating the resistive cathode and
pickup electrodes, both higher gaseous gain (more than $5\times 10^4$)
and strong spark reduction were achieved.
The spark probabilities have been checked using fast neutron
as a high ionizing particle (HIP). 
Those developments provide spark-free MPGD, which can be operated 
even in a next generation hadron collision experiment, 
where both MIP detection and HIP tolerance are required. 

\section{Design of detector}

\begin{figure}
\begin{center}
\includegraphics[height=8cm,bb=0 0 368 272]{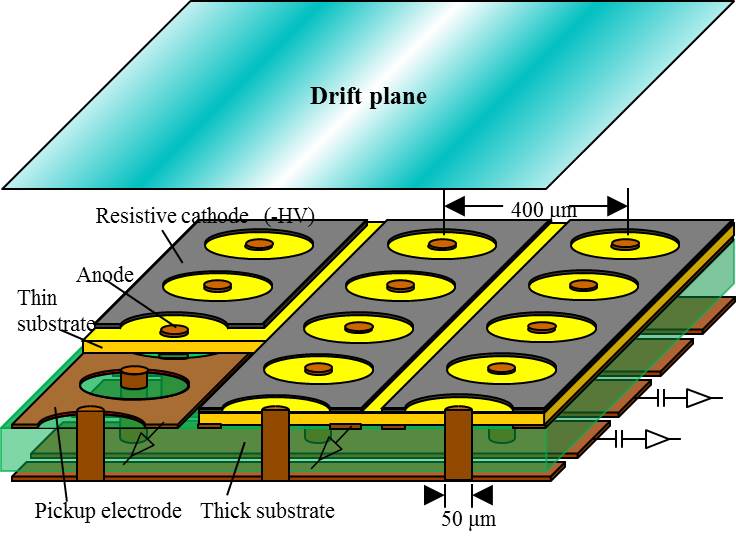}
\caption{\label{reupic_schematic} Schematic structure of $\mu$-PIC with
resistive cathode. Cathodes are made from resistive material (surface
resistivity is a few M$\Omega$/sq.). Signals from cathodes are read from
pickup electrodes as induced charge.}
\end{center}
\end{figure}

The structure of the new detector is based on $\mu$-PIC \cite{upic_1, upic_2}, 
which is one type of MPGD without floating structure, such as foils or mesh. 
Figure \ref{reupic_schematic} shows a schematic structure of resistive 
cathode $\mu$-PIC. 
The cathode electrodes are made from resistive material, 
instead of metal electrodes of existing $\mu$-PIC. 
Huge current caused by spark or large energy deposits will reduce 
the electric field by higher resistive cathodes. 
Signals are read from both anodes and pickup electrodes. 
Those two types of readouts cross each other at right angles 
for detecting two-dimensional position. 
The pickup electrodes are placed along resistive cathodes below an insulating layer. 
The signals are induced on the pickup electrodes 
with capacitive coupling on the detector surface.
In typical operation, static electrical field around the anode is 
quite the same as the existing $\mu$-PIC (cathodes are made from metal). 
Positive ions produced by gas multiplication drift toward the cathodes, 
and signals are induced on both anodes and pickup electrodes. 
Meanwhile, if there are large energy deposits or discharges, 
the electrical potential of the cathode is raised 
due to movement of a large amount of charge. 
It reduces the electric field around the anode, 
and terminates the discharge immediately. 
Maximum charge of a spark will be limited by very small 
(~0.1pF) capacitance of a single pixel.

\section{Prototype production}

\begin{figure}
\hspace{0.4cm}
\begin{minipage}{7cm}
\includegraphics[height=0.75\textheight,bb=0 0 364 905]{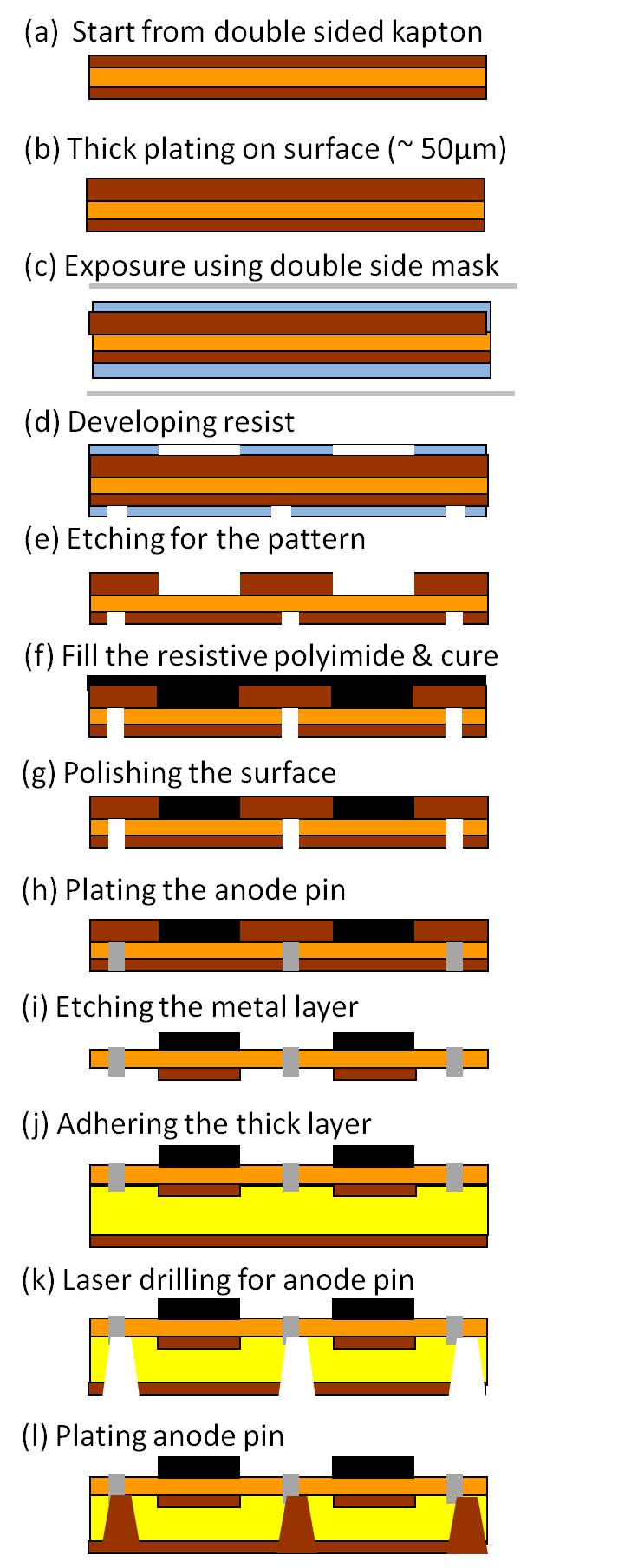}
\caption{\label{reupic_process} Manufacturing process of 
resistive $\mu$-PIC. Our prototypes are produced by RAYTECH Inc.}
\end{minipage}
\hspace{0.5mm}~
\begin{minipage}{8cm}
\includegraphics[width=8cm,bb=0 0 368 311]{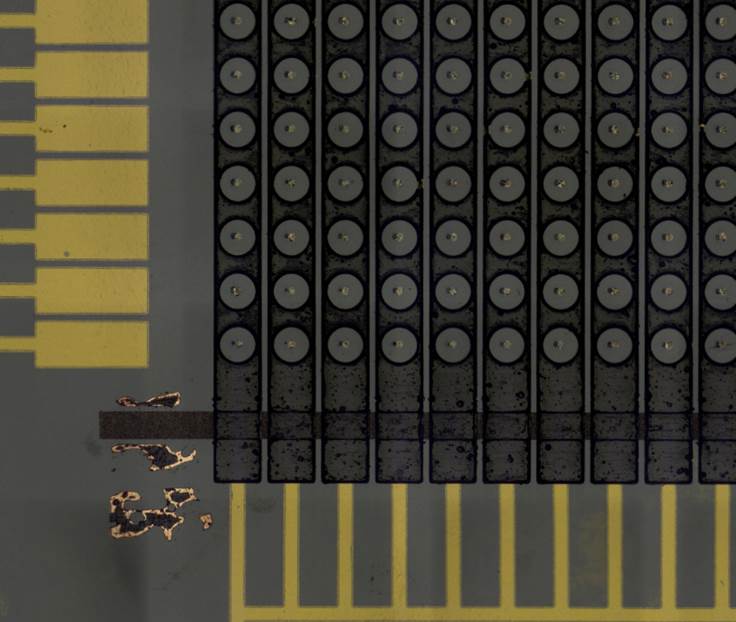}
\caption{\label{reupic_microscope} Microscopic picture of resistive $\mu$-PIC.
Strip pitch is 400$\mu$m. Potential of resistive strip is applied from
downside crossbar in this picture.}
\vspace*{0.5cm}
\includegraphics[width=8cm,bb=0 0 331 183]{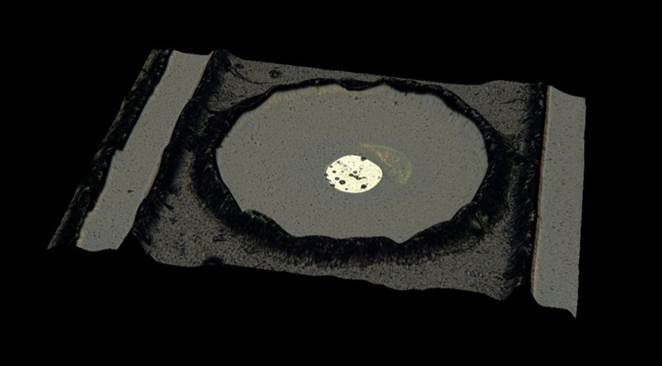}
\caption{\label{reupic_mag3D} Three-dimensional magnified image of one pixel.
3D image taken by digital microscope with focal depth information.}
\end{minipage}
\end{figure}

The prototype of the resistive cathode $\mu$-PIC was made 
using carbon-loaded polyimide, provided by Toray Industries, Inc. 
To make the resistive strips (cathode strips), 
images of surface cathode patterns were first etched on the PCB board. 
The carbon-loaded polyimide was laid on the surface, 
and remaining surface metal was removed chemically after curing the polyimide. 
Figure \ref{reupic_process} shows the manufacturing process of the
resistive cathode $\mu$-PIC.
Those manufacturing processes were done by Raytech Inc. 
Our prototype has 10cm $\times$ 10cm of detection area. 
The surface resistivity of the cathode strips is about 
100 k$\Omega$/sq. - 10M$\Omega$/sq. 
Figure \ref{reupic_microscope} shows 
a microscopic picture of the detector surface, 
and Figure \ref{reupic_mag3D} shows the magnified picture around one electrode.

\section{Operation test}

\begin{figure}
\begin{center}
\includegraphics[height=8cm,bb=0 0 517 388]{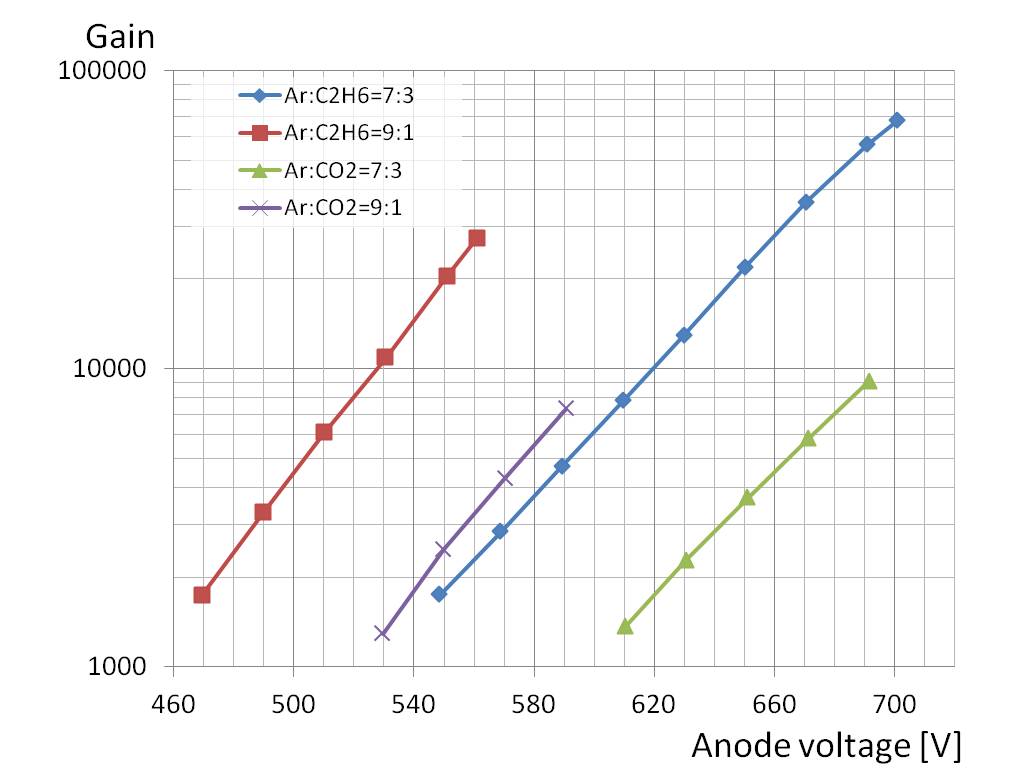}
\caption{\label{gaincurve} Gas gain of resistive $\mu$-PIC
against anode voltage. Potential of resistive cathodes was set to 0.
Four types of gas mixtures were tested: Argon/ethane = 7:3 and 9:1,
Argon/CO$_2$ = 7:3 and 9:1. Gain of more than 60,000 was attained
using Argon/ethane mixture.}
\end{center}
\end{figure}

The gaseous gain of the prototype was checked using $^{55}$Fe source
(5.9keV X-ray). 
Signals were observed from both anode and pickup strip. 
Signal lines were connected to ATLAS ASD \cite{asd} 
with analogue output as preamplifiers. 
The charges of the signals were measured by digital oscilloscope directly. 
The anodes and pickup signals have opposite signs 
but nearly identical charge in the same event. 
The gain curves were measured by the following four types of operation gases: 
Argon/ethane (7:3, 9:1) and Argon/CO$_2$ (7:3, 9:1).
A drift plane is set at a 3-mm gap from the $\mu$-PIC board while applying voltage
of -1 kV.
Potential of the resistive cathodes was set to 0 V, 
and the anode voltage was changed for gain measurements. 
The maximum anode voltage was determined by spark starting limit; 
However, the spark current of this resistive chamber (around 100 nA) was 
very low compared with conventional $\mu$-PIC (more than 1 $\mu$A).
Figure \ref{gaincurve} shows the results of gain measurements using those gases. 
The maximum gas gain was attained at $6\times 10^4$ using Argon/ethane (7:3) 
mixture gas.

\begin{figure}
\begin{center}
\includegraphics[width=11cm,bb=0 0 330 214]{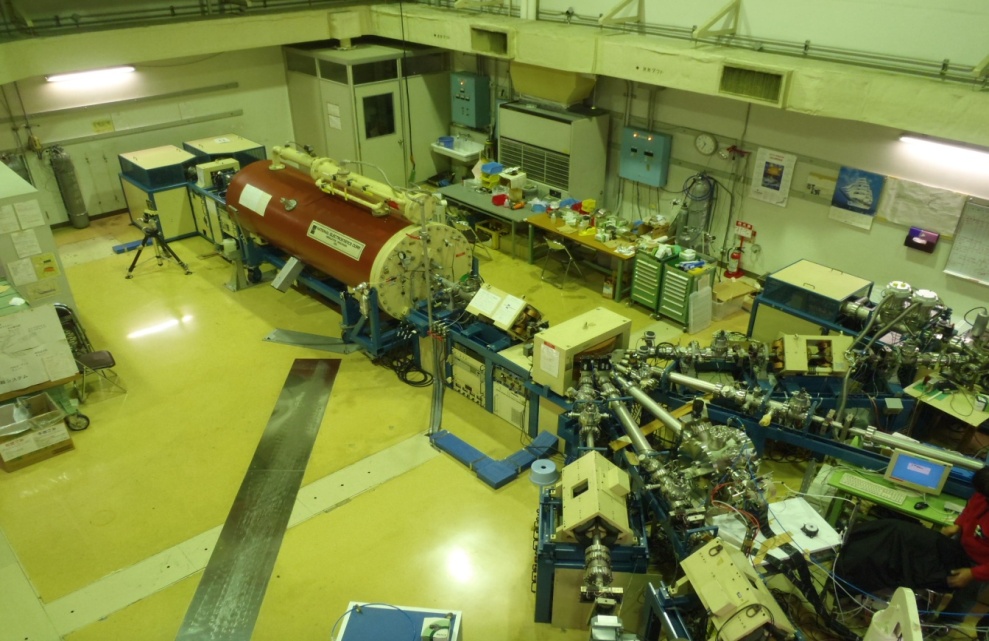}
\caption{\label{tandem_picture} Picture of a Tandem Electrostatic Accelerator 
at the Maritime Science Faculty, Kobe University. 
Ion sources are generated at the top of the system (upper-left part of picture), 
and accelerated in tandem accelerator. 
Maximum achievable energy (for proton or deuteron) is about 3.5 MeV.}
\end{center}
\end{figure}

\begin{figure}
\begin{center}
\includegraphics[height=8cm,bb=0 0 645 381]{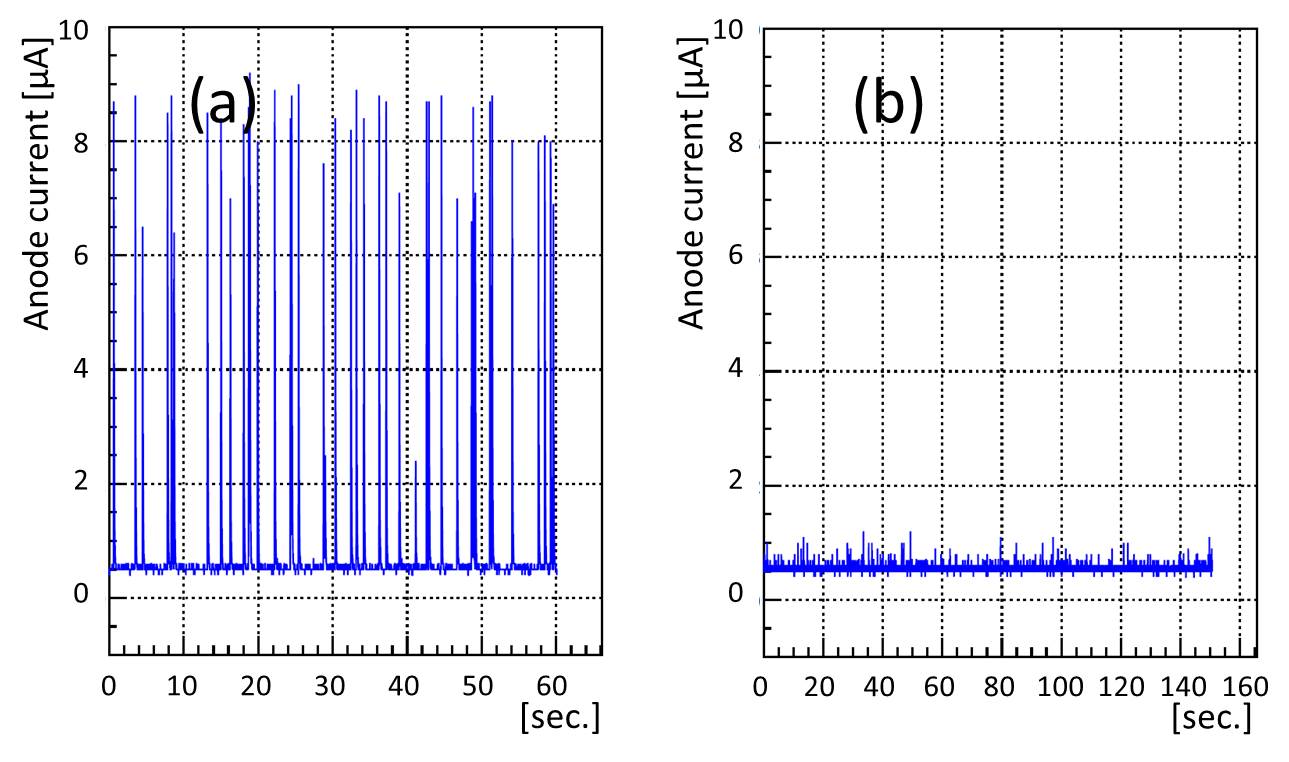}
\caption{\label{spark_current} Anode current on $\mu$-PIC
under intense neutron. (a) shows HV current using conventional $\mu$-PIC
with $2.4 \times 10^3$ neutron/second irradiation. (b) shows HV current
using resistive $\mu$-PIC with $1.9 \times 10^6$ neutron/second irradiation.
There is no spark found in case (b) at this scope; nevertheless it is $10^3$-times
more intense than case (a).}
\end{center}
\end{figure}

\begin{figure}
\begin{center}
\includegraphics[height=7cm,bb=0 0 455 376]{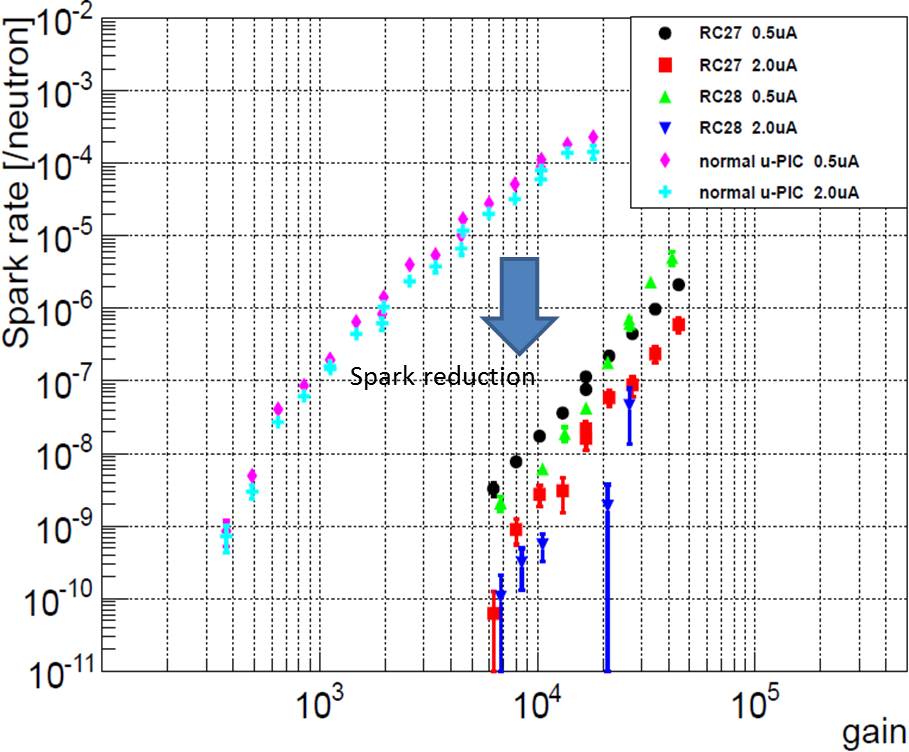}
\caption{\label{spark_rate}Spark rates of conventional $\mu$-PIC and 
resistive $\mu$-PICs.}
\end{center}
\end{figure}

The spark tolerances in HIP were measured using 
fast neutron irradiation on the chamber \cite{tipp2011_ochi}. 
A neutron source was provided by the tandem electrostatic accelerator 
at the Kobe University Faculty of Maritime Science (Fig. \ref{tandem_picture}). 
The 3-MeV deuteron beam generated by this accelerator was 
guided to a beryllium target placed at the end of the beam line. 
The neutron was produced by a $^9$Be(d,n)$^{10}$B 
exoergic (4 MeV) nuclear reaction. 
The peak neutron energy is estimated at 2.5 MeV. 
This facility can produce up to $10^8$ neutrons in a second. 
Figure \ref{spark_current}(a) shows the output current monitor 
for existing $\mu$-PIC with $2.4 \times 10^3$ neutron/second irradiation, 
and Figure \ref{spark_current}(b) shows resistive cathode 
$\mu$-PIC with $1.9 \times 10^6$ neutron/second irradiation. 
The gas gains of both setups were 15,000. 
We can see that the sparks were strongly suppressed, 
even with 900-times strong neutron.
The spark rate is plotted in Figure \ref{spark_rate}
with the gas gain on the abscissa. 
This rate is defined as the ratio of huge sparks to the irradiated neutron numbers. 
The sparks were counted when the current monitor of the anode 
HV excess was 0.5 $\mu$A and 2 $\mu$A. 
Two resistive cathodes $\mu$-PICs 
(named RC27/RC28 in the figure \ref{spark_rate}) 
and conventional $\mu$-PIC were tested. 
From those measurements, sparks in the resistive cathode $\mu$-PIC
were reduced to $10^3 \sim 10^5$-times smaller than those of conventional $\mu$-PIC.

\section{\label{novel_operation}
Novel Operation condition applying HV to resistive cathodes}

\begin{figure}
\begin{center}
\includegraphics[width=0.95\textwidth,bb=0 0 705 445]{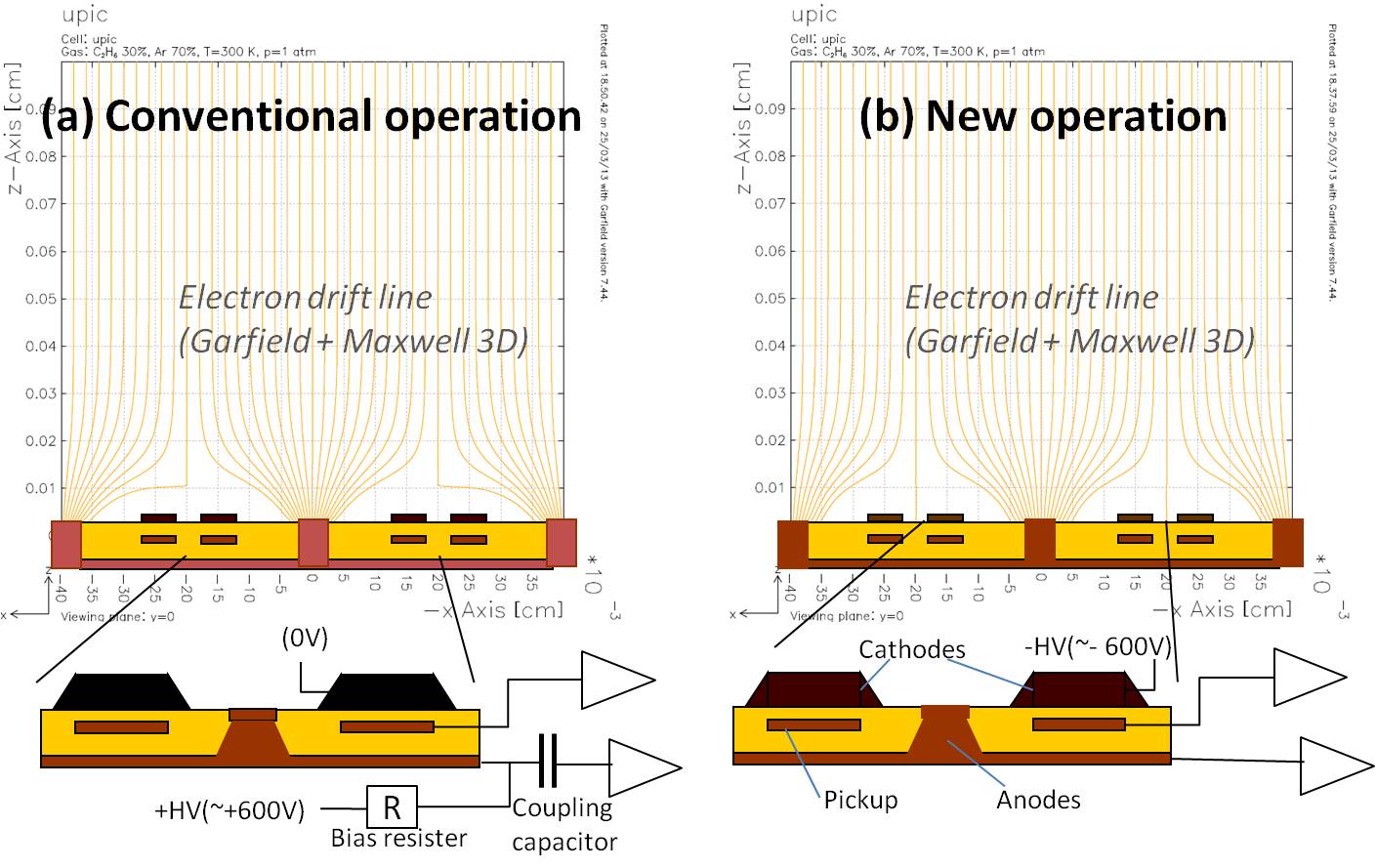}
\caption{\label{new_operation} Figure (a) shows electron drift line by applying
HV to resistive $\mu$-PIC as with conventional $\mu$-PIC. Positive HV
is applied to anodes, and potential of cathodes is set to 0 V. 
Figure (b) shows applying shifted HV on electrodes. Negative HV
is applied to cathodes, and anodes are connected to readout directly.
In (b), neither bias resisters nor coupling capacitors are needed.}
\end{center}
\end{figure}

In the conventional operation setup, positive HV was applied on anodes
and the potential of cathodes was set to 0 V, to make a high electric field
around the anode pixels. The coupling capacitors and
bias resisters were required for each anode strip to read signals. 
On the other hand, the potentials of pickup electrodes and cathodes 
can be independent in resistive $\mu$-PIC.
Then, shifting the surface potentials, in which the anode potential is set to 0 V
and resistive cathodes are applied negative voltage, is thought to work
almost the same as the conventional setup. 
Neither decoupling capacitors nor bias resisters are needed on anodes. 
Pickup electrodes, as second coordinate signals, can also be connected to
the readout directly. The capacitors and resisters for each electrode can be
completely removed using this new operation. Difference between conventional
and new operation modes is shown in Figure \ref{new_operation} with electron
drift lines simulated using Garfield and Maxwell 3D.
Both electron drift lines seem almost the same in those simulations.

\begin{figure}
\hspace{0.4cm}
\begin{minipage}{7cm}
\includegraphics[width=7cm,bb=0 0 270 206]{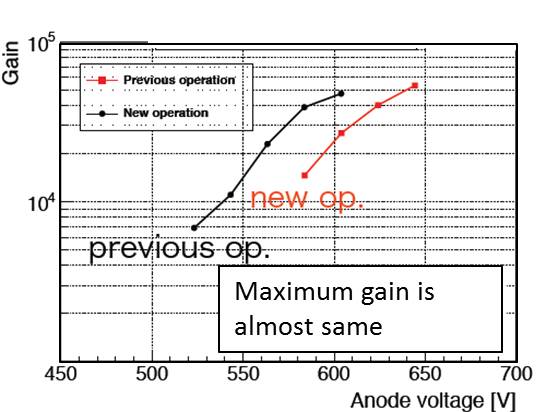}
\caption{\label{new_gaincurve} Gain curve of resistive $\mu$-PIC with
conventional operation and new operation. $^55$Fe (5.9keV) source was
used with Argon/ethan (7:3) gas mixture.}
\end{minipage}
\hspace{0.5mm}~
\begin{minipage}{8cm}
\includegraphics[width=8cm,bb=0 0 425 290]{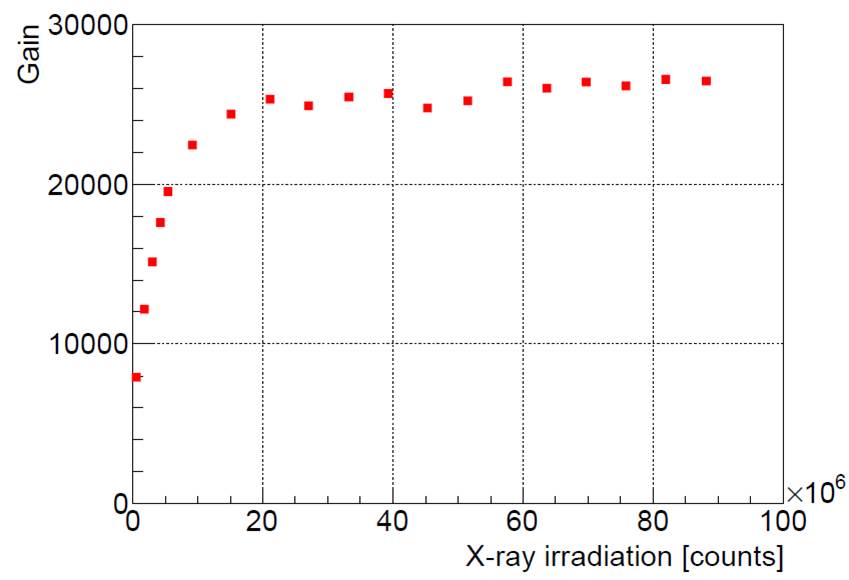}
\caption{\label{time_val} Time variability of gain on new operation mode.
Gas gain of 8,000 at start grew to 25,000 after $2\times 10^7 $counts/cm$^2$
X-ray (5.9keV) irradiation.}
\end{minipage}
\end{figure}

\begin{figure}
\begin{center}
\includegraphics[width=9cm,bb=0 0 316 230]{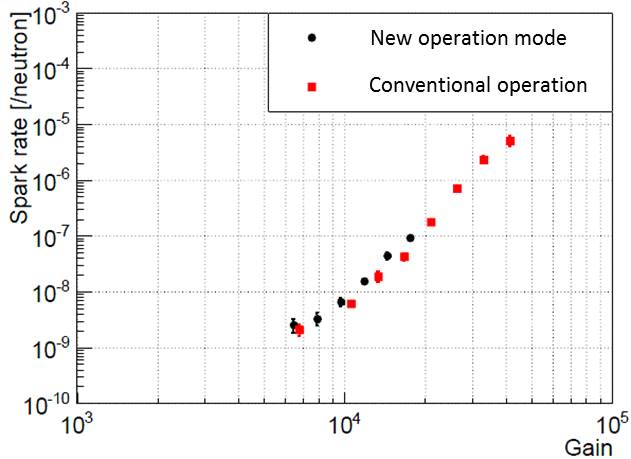}
\caption{\label{new_spark} Spark probabilities of resistive $\mu$-PIC with
both conventional and new operation under fast neutron irradiation.}
\end{center}
\end{figure}

The gain curve and spark probabilities were measured in the new operation
condition. The gain curve using Argon/ethane (7:3) mixture gas is shown
in Figure \ref{new_gaincurve}. Operation voltage with the new operation was
a bit greater than that with the conventional one. 
The time variability of the gain was found as shown in Figure \ref{time_val}.
It is thought that this phenomenon is due to charge-up on the substrate.
The gain became stable after $2\times 10^7$counts/cm$^2$ irradiation
of incident 5.9keV ($^{55}$Fe) X-ray source.
The maximum gain and shape of gain curves were almost the same.
As shown in Figure \ref{new_spark}, spark probabilities under fast neutron
were also almost the same.

\section{Discussion and future prospects}

\begin{figure}
\hspace{0.4cm}
\begin{minipage}{8cm}
\includegraphics[width=8cm,bb=0 0 429 303]{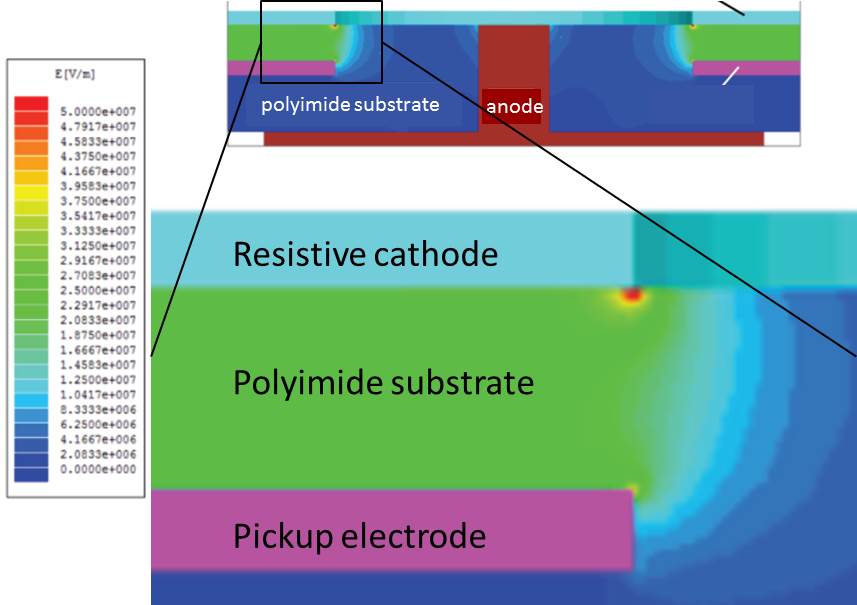}
\caption{\label{field_edge} Electric field around cathode and pickup electrodes,
simulated using Maxwell 3D.
High electric field is found at edge of resistive cathode. }
\end{minipage}
\hspace{0.5mm}~
\begin{minipage}{7cm}
\includegraphics[width=7cm,bb=0 0 222 163]{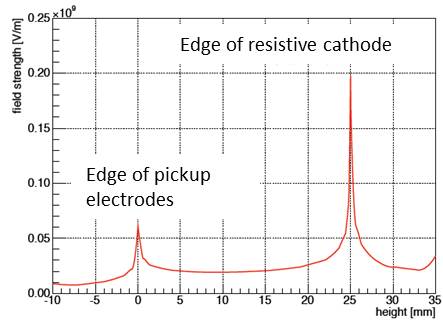}
\caption{\label{field_xsection} Field strength along line joining both edges
of cathode and pickup electrode.}
\end{minipage}
\end{figure}

The results from section \ref{novel_operation} show the new operation condition
is performed as well as the conventional one. The new one has a great advantage
in terms of detector production because fewer electronic parts are needed.
However, there are problems remaining for stable operation.
Sometimes the pickup electrodes have been shorted to resistive cathodes.
Two of three of our prototypes have been broken in this phenomenon.
The withstanding electric field of polyimide is around 300 kV/mm,
and the substrate thickness (between resistive cathode and pickup) is 25$\mu$m.
The average electrical field in the substrate is estimated at around 20 kV/mm,
which seems sufficiently lower than the withstanding field of polyimide.
However, it is expected that the field at the edge of the electrodes will be higher.
The electric fields around the cathode and pickup edge are simulated using
Maxwell 3D as shown in Figure \ref{field_edge}.
Figure \ref{field_xsection} shows the field strength along the line joining
both the cathode and the pickup edge in Figure \ref{field_edge}.
The maximum field is about 200 kV/mm at the cathode edge, which is
close to the withstanding field of polyimide. 
This simulation suggests that the margin of withstanding voltage is not sufficient.
To improve it, thicker substrate will be used in the next prototype.

\section{Conclusion}
The Micro Pixel Chamber with resistive cathodes and capacitive readout
was developed and tested. More than $6\times 10^4$ of gas gain was
achieved stably, using $^{55}$Fe source under Argon/ethane (7:3) gas mixture.
The spark rates were also measured using intense fast (~2MeV) neutrons.
The spark rate of resistive $\mu$-PIC was strongly reduced.
It was $10^4$-times smaller than that of conventional $\mu$-PIC at
gain of $10^4$.
Using capacitive readout, the new operation mode, in which 
the two-dimensional signals can be read without HV coupling, was achieved.
Although we need more improvements in the production process for stable operation,
the resistive $\mu$-PIC will be one of the best solutions for non-floating
structure MPGD.

\section*{Acknowledgments}
We thank Hideo Uehara in Reytech-Inc. whose diligent efforts contributed
to our proposal and improved the fabrication process. 
Also we thank the RD51 members, especially Rui de Oliveira
for many fruitful suggestions,
and we also would like to thank the KEK Detector Technology Project
and the Japan MPGD Basic R\&D Team for their suggestions and technical supports.
This research was supported by Grant-in-Aid for Scientific Research 
(No.23340072, 24654067).

\end{document}